\documentstyle[epsfig]{article}
\begin{document}

\author{{\bf Jo\~{a}o Martins}\footnote{%
Laborat\'{o}rio de Mecatr\'{o}nica, DEEC, IST, Av. Rovisco Pais, 1096 Lisboa
Codex, Portugal}  \footnote{%
Escola Superior de Tecnologia, IPS, Rua do Vale de Chaves, Estefenilha, 2910
Set\'{u}bal, Portugal} and {\bf R. Vilela Mendes}\thanks{%
Grupo de F\'{i}sica-Matem\'{a}tica, Complexo Interdisciplinar, Univ. de
Lisboa, Av. Gama Pinto 2, 1699 Lisboa Codex, Portugal}{
\footnotemark[1] }}
\title{Neural networks and logical reasoning systems. A translation table }
\date{}
\maketitle

\begin{abstract}
A correspondence is established between the elements of logic reasoning
systems (knowledge bases, rules, inference and queries) and the hardware and
dynamical operations of neural networks. The correspondence is framed as a
general translation dictionary which, hopefully, will allow to go back and
forth between symbolic and network formulations, a desirable step in
learning-oriented systems and multicomputer networks.

In the framework of Horn clause logics it is found that atomic propositions
with n arguments correspond to nodes with n-th order synapses, rules to
synaptic intensity constraints, forward chaining to synaptic dynamics and
queries either to simple node activation or to a query tensor dynamics.
\end{abstract}

\section{Introduction}

Neural networks and logical reasoning systems, because they both rely on
Turing models of computation, are equivalent in the sense that whatever is
computable in one framework must also be computable in the other. How to
establish the equivalence in each particular case is however a non-trivial
and interesting issue, because problems that are addressed in a simple
manner in one approach turn out to be intractable in the other and
vice-versa. For example, combinatorial optimization problems are efficiently
handled by neural networks, whereas logical systems succumb to combinatorial
explosion. Conversely, for highly structured problems where appropriate
heuristics are known, it is simpler and faster to use a logical reasoning
system, the corresponding neural network system taking a long time to tune
up because rule-based information is in general not so easy to build in the
network architecture.

This situation led a number of authors\cite{Barnden1} \cite{Honavar} to
propose the use of hybrid systems to solve complex problems, with a
logic-like program for some parts of the process and a network for the rest.
On the other hand, some authors have dealt with the identification of the
network structures that might represent the logical structures of a symbolic
system\cite{Shastri} \cite{Pollack} \cite{Pinker} \cite{Barnden2} \cite{Sun1}
\cite{Sun2} \cite{Sethi} \cite{Botta}and conversely with the nature of the
rule-based information that may be extracted from networks.

Extraction of rules from networks is an issue of practical importance in the
construction of expert systems from example-trained networks\cite{Gallant1} 
\cite{Gallant2} \cite{Towell} \cite{Fu}. On the other hand when some prior
rule-based information is known about a problem but nevertheless a network
implementation seems appropriate, it would be useful to have some simple
rules to implement the symbolic information on the architecture of the
network. Neural networks take advantage of parallel VLSI hardware
implementations, which largely improve the processing speed and it is not so
clear how to take advantage of similar implementations for symbolic
processing. Therefore even for problems that are naturally stated in logic
terms, it might be useful to have a translation dictionary for hardware
implementation purposes. The establishment of a concise way to go back and
forth between symbolic and network formulations is also welcome in
learning-oriented systems and in the design of multicomputer networks.
Different architectures make different types of learning easier or harder to
design and in multicomputer networks it is essential for algorithms and
architectures to fit together as well as possible. Finally and independently
of the practical issues, the establishing of a compact translation
dictionary between the two paradigms might be a useful step in the
development of a unified language for cognitive processes.

Most past attempts at establishing a logic-networks dictionary concern
questions of the architecture required to represent some types of logical
operations or the refining of the numerical part of the knowledge base. In
some cases, an extension of the usual connectionist framework is required
and a full correspondence is in general not established involving all the
aspects of the systems (knowledge base, rules, inference, recursion and
handling of queries). Here we describe an attempt to establish a
correspondence involving all the basic elements that are present in logical
and neural network systems. Recall that in a neural network one has a
(distributed) memory on the connection strengths (synapses), a learning
dynamics on the synapses and an activation dynamics of the node intensities.
In a logical reasoning system there is some set of ground facts about the
objects in the domain, a set of rules which are potential knowledge
concerning relations between the objects and an inference mechanism
(backward or forward chaining) allowing for the extraction of further
information and the answering of queries.

It has been argued that trying to isolate, in a network, the structures that
correspond to specific logical statements or operations is a waste of time
because everything in a network (memory, rules and inference) is distributed
everywhere and forever inseparable. This may well be true for some
architectures and some classes of concepts. However even the identification
of the modular structures that correspond to the logic elements is already a
useful step. For example one of our conclusions here is that a natural
network representation of an atomic proposition is a node with n$^{th}$
order synapses. Because of the universal approximation properties of neural
networks, that same proposition might as well be represented with first
order synapses, the proposition corresponding then to a submodule of several
neurons. However the identification of the kind of minimal submodule that
corresponds to a specific logical element is already a useful step.

\section{The network translation of a logic system}

For definiteness the scope of the logical systems, that is considered is a
subset of the Prolog logic programming language. Namely the logic system is
specified by a set of {\it Horn clauses} which are constructed from {\it %
terms} which are either {\it constants} or {\it variables} or {\it structures%
}. A constant is sensed to represent some concrete object in the domain of
the problem. It is represented in the logical system by an indecomposable
elementary symbol (an {\it atom}). Structures are restricted to be atomic
propositions of the general form 
\[
\textnormal{functor(parameter list)} 
\]
The functor is an atom and the parameter list is any list of atoms,
variables or other atomic propositions. Finally a variable is an entity that
can at any time be bound to any atom (constant or functor). Small letters
will be used for the atoms and capitals for the variables.

The first step is to find a network representation of the basic entities of
the logical system. Each atom will correspond to a {\it node}. Therefore not
only constants but also functors will be assigned a node in the network.

The network nodes have output zero or one and threshold one. The nodes are
connected in a network, with the inputs equipped with connection strengths
(synapses) $W_{f,i_{1},\cdots ,i_{n}}$ of order $n$\cite{Softy} \cite{Taylor}
\cite{Clarke} if $n$ is the number of parameters in the parameter list of
the functor. In particular a simple constant has no input. The node
activation is 
\begin{equation}
x_{f}=\theta \left( \sum_{i_{1}\cdots i_{n}}W_{f,i_{1},\cdots
,i_{n}}x_{1}\cdots x_{n}+y_{f}\right)  \label{4.1}
\end{equation}
where the polarization $y_{f}$ is zero unless there is a query involving
this node. Fig. 1 displays the neural network representation of a {\it %
constant node} and Fig. 2 that of a {\it functor node.}

Whenever in the knowledge base, the relation 
\begin{equation}
f(i_{1},\cdots ,i_{n}).  \label{4.2}
\end{equation}
is stated as a true fact, the corresponding synaptic strength is 
\begin{equation}
W_{f,i_{1},\cdots ,i_{n}}=1  \label{4.3}
\end{equation}
\begin{figure}[htb]
\begin{center}
\psfig{figure=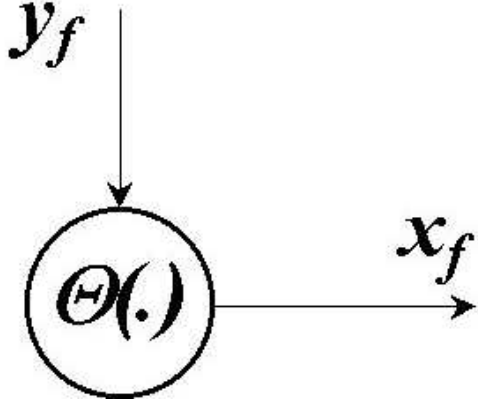,width=7truecm}
\caption []{Network representation of a constant}
\end{center}
\end{figure}
\begin{figure}[htb]
\begin{center}
\psfig{figure=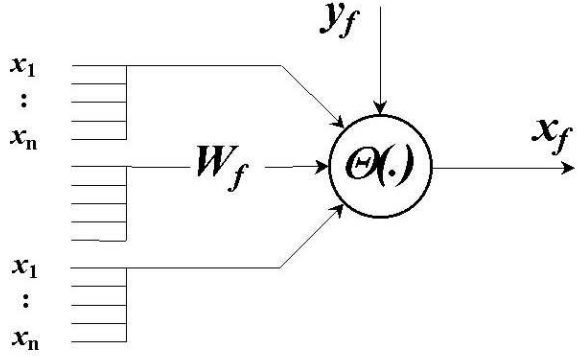,width=8truecm}
\caption []{Network representation of a functor}
\end{center}
\end{figure}

Otherwise, if the relation $f$ does not hold for these particular set of
parameters the connection strength vanishes. This is the straightforward way
to represent in the network the {\it ground facts} of the knowledge base,
that is, the relations that are explicitly expressed among the atoms. On the
other hand relations and clauses that involve variables correspond to
constraints on the remaining connection strengths. This is better explained
through an example. Let a knowledge base be 
\begin{equation}
\begin{array}{ccl}
f(b,c). &  &  \\ 
m(a,c). &  &  \\ 
m(d,a). &  &  \\ 
p(X,Y) & \Leftarrow & m(X,Y). \\ 
p(X,Y) & \Leftarrow & f(X,Y). \\ 
g(X,Z) & \Leftarrow & p(X,Y)\wedge p(Y,Z).
\end{array}
\label{4.4}
\end{equation}
The ground facts (the first three headless clauses) imply 
\begin{equation}
W_{fbc}=1;W_{mac}=1;W_{mda}=1  \label{4.6}
\end{equation}
all other $W_{fXY}$ and $W_{mXY}$ being set to zero. Let the connection
strengths take only the values zero and one. Then the last three clauses in (%
\ref{4.4}) translate into the following constraints on the remaining $%
W_{pXY} $ and $W_{gXY}$ connection strengths 
\begin{equation}
\begin{array}{ccl}
W_{pXY} & \geq & W_{mXY} \\ 
W_{pXY} & \geq & W_{fXY} \\ 
W_{gXZ} & \geq & W_{pXY}W_{pYZ}
\end{array}
\label{4.7}
\end{equation}
These constraints may be realized in the network by a parallel dynamical
law. Define the potential function 
\begin{equation}
V_{1}=\frac{1}{2}\sum_{XY}\chi \left( W_{mXY}-W_{pXY}\right) +\frac{1}{2}%
\sum_{XY}\chi \left( W_{fXY}-W_{pXY}\right) +\frac{1}{2}\sum_{XYZ}\chi
\left( W_{pXY}W_{pYZ}-W_{gXZ}\right)  \label{4.8}
\end{equation}
where $\chi (x)=x^{2}\theta (x)$. Now the strengths (\ref{4.6}) fixed by the
ground statements are considered as frozen and the others evolve in parallel
according to a gradient descent dynamics 
\begin{equation}
\frac{dW_{pXY}}{dt}=-\frac{\partial V_{1}}{\partial W_{pXY}}\quad ;\quad 
\frac{dW_{gXY}}{dt}=-\frac{\partial V_{1}}{\partial W_{gXY}}  \label{4.9}
\end{equation}
This dynamics drives the connection strengths to values zero or one
satisfying the constraints (\ref{4.7}). Fig. 3 shows the dynamical evolution
of the connection strenghts for the potential (\ref{4.8}).
\begin{figure}[htb]
\begin{center}
\psfig{figure=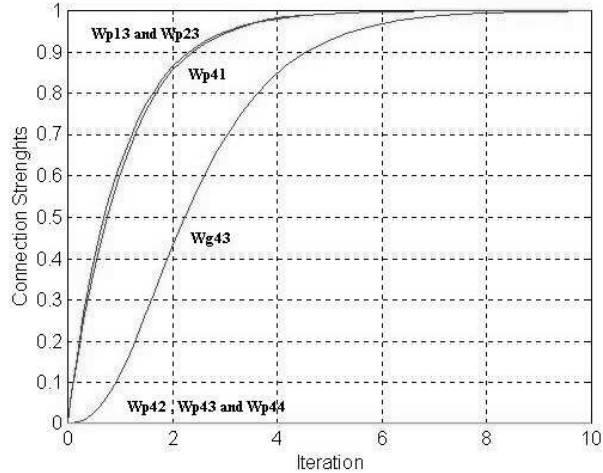,width=9truecm}
\caption []{A connection strengths evolution example}
\end{center}
\end{figure}

The method used in this example holds in general and for each rule of the
form 
\begin{equation}
g\left( X_{1},...,X_{n}\right) \Leftarrow \prod_{i}\wedge f_{i}\left(
Y_{1},...,Y_{p_{i}}\right)  \label{4.10}
\end{equation}
a term 
\begin{equation}
\frac{1}{2}\sum_{\{X\}\{Y\}}\chi \left(
\prod_{i}W_{f_{i}Y_{1}...Y_{p_{i}}}-W_{gX_{1}...X_{n}}\right)  \label{4.11}
\end{equation}
is added to the potential $V_{1}$ that drives the connection strengths, the
sum being over all the variable terms.

We have therefore established a correspondence where atomic propositions
with $n$ parameters correspond to $n^{th}$ order nodes and the rules are
implemented by a parallel dynamics of the synapses. Because the synapse
dynamics is equivalent to a forward chaining step, the network now contains
on its connection strengths all the available information about the problem.
Therefore information may be retrieved by simply looking at the values of
the connection strengths. Alternatively an explicit scheme to answer queries
may be established in the following way.

We may consider two different types of queries, simple queries without
variables and queries with variables. In the first case, for example 
\begin{equation}
?\Leftarrow g(a,b).  \label{4.12}
\end{equation}
the question is to check whether the relation $g$ holds for the
atoms $a$ and $b$. In this case it suffices to excite the polarizations $%
y_{a}$ and $y_{b}$ of the nodes $a$ and $b$ and see whether the node $g$
lights up. The same happens for more complex queries like, for example 
\begin{equation}
?\Leftarrow g(h(a,b),c)\wedge f(c,d).  \label{4.13}
\end{equation}
where now the polarizations to be excited are $y_{a}$, $y_{b}$, $y_{c}$ and $%
y_{d}$ and the answer to the query is positive if nodes $g$ and $f$ light up
simultaneously.

For queries involving variables, for example 
\begin{equation}
?\Leftarrow m(d,X)\wedge p(X,c).  \label{4.14}
\end{equation}
the question is to find the instantiations of $X$, if any, that make this a
true statement. The product of the connection strengths $W_{mdX}W_{pXc}=%
\alpha _{X}$ defines a vector $\alpha _{X}$ in atom space and the non-zero
entries of the vector are the instantiations that positively satisfy the
query. A similar scheme holds for queries involving several variables. For
example, for 
\begin{equation}
?\Leftarrow g(h(X,c),a)\wedge f(c,Y).  \label{4.15}
\end{equation}
ones defines a tensor $\beta _{XY}=W_{gha}W_{hXc}W_{fcY}$ , the non-zeros
entries of which are the positive answers to the query. With all the
connection strengths established in the network by the (forward
chaining-like) dynamics of Eq.(\ref{4.9}) it might be an easy matter
(depending on the type of implementation used) to look up the non-zero
entries of $\alpha _{X}$ or $\beta _{XY}$ . Alternatively one may regard the
network connection strengths as potential knowledge and generate dynamically
the non-zero entries of the tensors 
\begin{equation}
\frac{d\alpha _{X}}{dt}=-\frac{\partial V_{2}}{\partial \alpha _{X}}\quad
;\quad \frac{d\beta _{XY}}{dt}=-\frac{\partial V_{3}}{\partial \beta _{XY}}
\label{4.16}
\end{equation}
with 
\begin{equation}
\begin{array}{lll}
V_{2} & = & \sum_{X}\left( W_{mdX}W_{pXc}-\alpha _{X}\right) ^{2} \\ 
V_{3} & = & \sum_{XY}\left( W_{gha}W_{hXc}W_{fcY}-\beta _{XY}\right) ^{2}
\end{array}
\label{4.17}
\end{equation}
Regarding the network structure and connection strengths as potential, as
opposite to explicit knowledge, the dynamics of the query tensors defined
above plays a role similar to backward chaining in logical programming.

As in (\ref{4.10})-(\ref{4.11}) the generalization, of this dynamics in the
space of query tensors, to general queries is straightforward.

The whole construction is summarized in Table 1.

\smallskip

\begin{tabular}{|l|l|}
\hline
{\bf Logical System} & {\bf Network} \\ \hline
Constant & Single node \\ \hline
Atomic proposition ($n$ arguments) & Node with $n^{th}$-order synapses \\ 
\hline
Rules & Synaptic intensity constrains \\ \hline
Rule application (forward chaining) & Synaptic dynamics \\ \hline
Queries & Node activation / Query tensor dynamics \\ \hline
\end{tabular}

\begin{center}
Table 1: Logical system - Neural network translation table
\end{center}

As a simple example we discuss {\bf problem 148} of Ref. \cite{Coelho} (see
Appendix). Let us consider the following ground facts:

\begin{center}
\begin{tabular}{l}
{\it phone(coleman,'100001').} \\ 
{\it phone(gordon,'100002').} \\ 
{\it phone(wagner,'100003').} \\ 
{\it phone(smith,'100004').} \\ 
{\it visits(coleman,wagner).} \\ 
{\it at(wagner,gordon).}
\end{tabular}
\end{center}

The clauses in this example define the following potential 
\begin{equation}
V_{4}=\frac{1}{2}\sum_{XYZ}\chi \left( W_{visitsXY}W_{atYZ}-W_{atXZ}\right) +%
\frac{1}{2}\sum_{UVN}\chi \left( W_{atUV}W_{phoneVN}-W_{numberUN}\right)
\label{4.18}
\end{equation}
.

The evolution of the connection strengths, according to the gradient
descendent dynamics, leads to the neural network in Fig. 4, where only the
non-zero connections strengths are shown. The continuous lines represent the
connection strenghts of the ground facts and the dashed ones the connection
strenghts obtained from the clauses used to construct the potential $V_{4}$.
\begin{figure}[htb]
\begin{center}
\psfig{figure=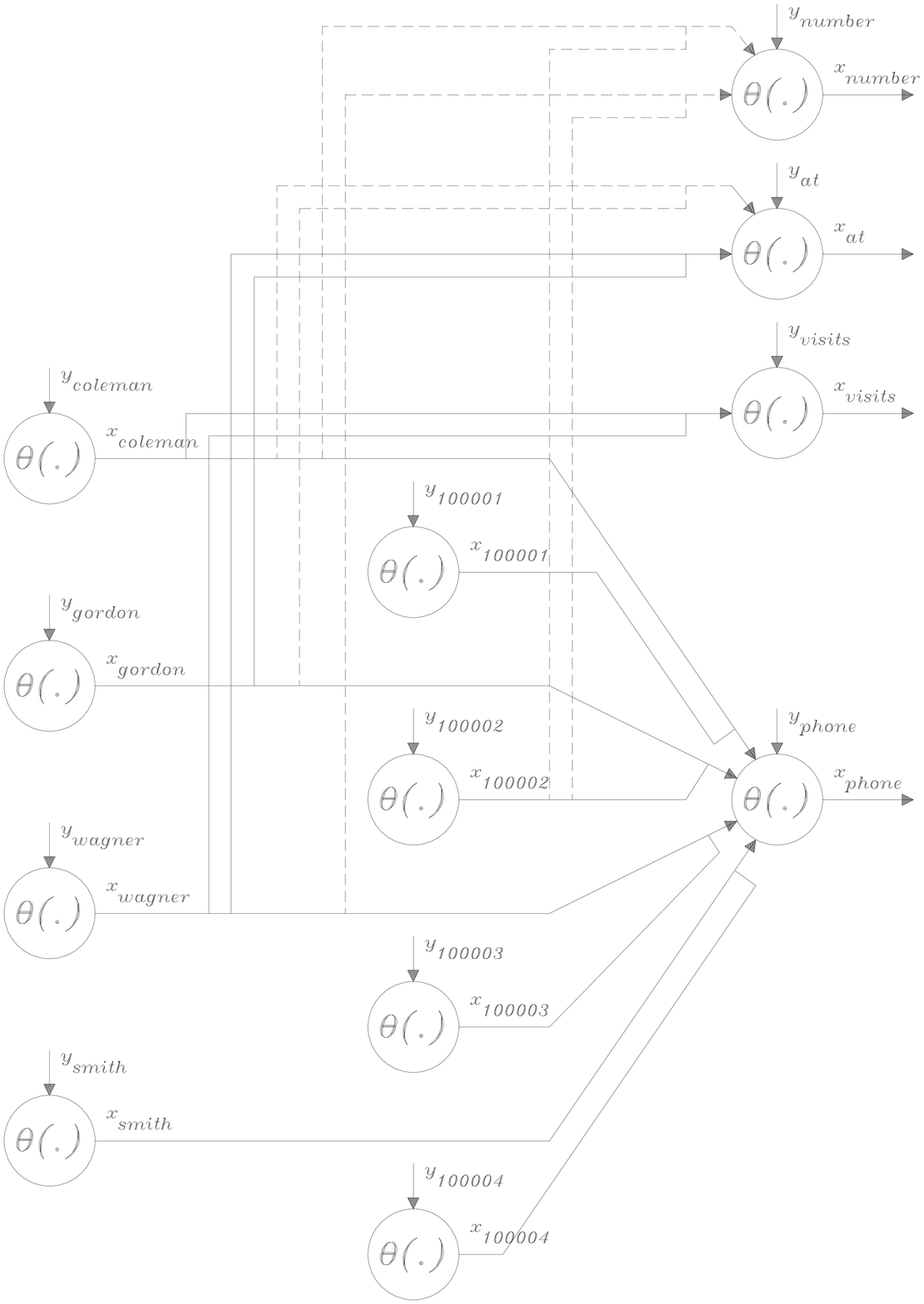,width=11truecm}
\caption []{Neural network representation of problem 148 in Ref.\cite{Coelho}}
\end{center}
\end{figure}

\section{Conclusion and remarks}

The construction in the previous section emphasized the establishment of a
correspondence between the pure logic elements of logic programming and both
the hardware and dynamical operations of neural networks. Logic programming
systems however, in addition to its pure logic elements, involve some
non-logical features like the handling of arithmetics and procedural
prescriptions for computing efficiency. These features need not concern us
here because numerical capabilities are native to networks and procedural
rules are implementation dependent.

A more serious limitation concerns the restriction to Horn clauses. Horn
clause logic is quite expressive and of wide computing power. However it
cannot handle negation in a satisfactory way, especially for non-exhaustive
databases. In some cases it is acceptable to reduce, by renaming\cite{Nie},
the set of non-Horn to a Horn clause set. However, depending on the nature
of the database, it may be more convenient to retain a certain degree of
non-Horness and to establish an inference procedure through case analysis%
\cite{Loveland} \cite{Plaisted1} \cite{Plaisted2}. At a formal level, and to
some extent, our formalism might be extended to non-Horn clauses. For
example, if a non-Horn clause is 
\begin{equation}
h(X,Z)\vee g(X,Z)\Leftarrow p(X,Y)\wedge p(Y,Z).  \label{4.19}
\end{equation}
the relation for the connection strengths that extends (\ref{4.7}) is 
\begin{equation}
W_{hXZ}+W_{gXZ}\geq W_{pXY}W_{pYZ}
\end{equation}
However it is not clear, at this time, how to implement dynamically, in an
elegant manner, the case analysis mechanism.

In this paper we have been concerned with logical reasoning systems in its
sharp logics version. A natural concern is the extension to reasoning with
uncertainty. Given that the handling of real numbers is quite natural in
networks, an extension where all quantities (node activations and connection
strengths), instead of being $0$ or $1$, take a continuous range of values,
does not seem to raise any special difficulties. Notice however that the
straightforward extension of our network operations, as defined in Sect. 2,
lead to represent the fuzzy intersection by a product rather than by the $%
\min $ operation.

\section{Appendix}

{\bf Problem 184} in Ref. \cite{Coelho}, originally suggested in ''The
Thinking Computer'' by Bertrand Raphael, considers the following functors:

\begin{center}
\begin{tabular}{ll}
{\it number(X,N)} & means that person X can be reached by calling phone
number N. \\ 
{\it visits(X,Y)} & means that person X is visiting person Y. \\ 
{\it at(X,Y)} & means that person X is at the residence of person Y. \\ 
{\it phone(X,N)} & means that person X has phone number N.
\end{tabular}
\end{center}

and the following clauses:

\begin{center}
\begin{tabular}{lll}
{\it at(X,Z)} & $\Leftarrow $ & {\it visits(X,Y) }$\wedge ${\it \ at(Y,Z)}
\\ 
{\it number(U,N)} & $\Leftarrow $ & {\it at(U,V) }$\wedge ${\it \ phone(V,N)}
\end{tabular}
\end{center}

The purpose is to reach a person, having a phone number data base and
knowing who visits whom and where the visited person is.

\end{document}